\documentclass[]{emulateapj}
\usepackage{psfig}
\submitted{Accepted to The Astrophysical Journal, Letters (December 16, 2003)}

\advance \voffset by 0.00cm\relax
\def\msun{{\rm ~M}_{\odot}}

\shorttitle{X-RAY BINARY POPULATIONS: NGC1569}
\shortauthors{BELCZYNSKI, KALOGERA, ZEZAS, \& FABBIANO }

\begin{document}

\title{X-RAY BINARY POPULATIONS: THE LUMINOSITY FUNCTION OF NGC1569}

\author{K.\ Belczynski\altaffilmark{1,2}, V.\ Kalogera\altaffilmark{1}, 
        A.\ Zezas \altaffilmark{3}, and G.\ Fabbiano\altaffilmark{3}}

\affil{ $^{1}$ Northwestern University, Physics \& Astronomy,
       2145 Sheridan Rd., Evanston, IL 60208\\ 
       $^{2}$ Lindheimer Postdoctoral Fellow\\
       $^{3}$ Harvard-Smithsonian Center for Astrophysics, 60 Garden St., Cambridge, 
       MA 02138\\
       belczynski, vicky@northwestern.edu; azezas, pepi@head.cfa.harvard.edu}

 \begin{abstract} 
 Using the population synthesis code {\em StarTrack} we construct the first
 synthetic X-ray binary populations for direct comparison with the X-ray 
 luminosity function (XLF) of NGC~1569 observed with {\em Chandra}. 
 Our main goal is to examine whether it is possible to reproduce the XLF 
 shape with our models, given the current knowledge for the star-formation 
 history of this starburst galaxy. We thus produce hybrid models meant 
 to represent the two stellar populations: 
 one old, metal-poor with continuous star-formation for $\sim$ 1.5\,Gyr and 
 another recent and metal-rich population. To examine the validity of the models 
 we compare XLFs calculated for varying ages of the populations and varying 
 relative weights for the star-formation rates in the two populations. 
 We find that, for typical binary evolution parameters, it is 
 indeed possible to quite closely match the observed XLF shape. 
 The robust match is achieved for an age of
 the young population and a ratio of star formation rates in the two
 populations that are within factors of 1.5 and 2, respectively, of those 
 inferred from HST observations of NGC~1569. 
 In view of this encouraging first step, we discuss the implications of our X-ray
 binary models and their potential as tools to study binary populations in 
 galaxies.

 \end{abstract}


\section{INTRODUCTION}
\label{sec:intro}

X-ray binaries (XRB) in nearby galaxies  have been known since 
the {\em Einstein} era, but detections of significant samples were limited to 
just a few sources
(e.g., Magellanic Clouds or M31) and interpretation of the source properties was 
often hampered by confusion problems. {\em Chandra} observations 
have revolutionized XRB studies with the discovery of 
large numbers of point X-ray sources in galaxies even beyond the Local 
Group (e.g., Fabbiano \& White 2003). Short-term variability of many sources  
excludes the possibility of source confusion and 
strongly points toward accretion as the origin of the X-rays. 
The samples in most cases are large enough that we are now able to examine XRB 
{\em populations} in a wide range of galactic environments with different star 
formation histories. The samples are characterized 
by cumulative X-ray luminosity functions (XLFs) fitted by single or broken 
power-laws (e.g., Grimm et al.\  2003; Zezas \& Fabbiano 2002). 

It has been noted that the XLF slopes follow a rather systematic behavior with  
population age and possibly star formation rate (SFR) (Grimm et al. 2003; 
Sarazin et al. 2003). 
Such  behavior could constrain the SF history 
and properties of XRB populations in nearby galaxies. However, the 
development of  reliable diagnostics requires a sound physical understanding of the 
observed correlations. With the exception of a recent study (Sipior et al. 2003, 
which does not focus on a specific set of observations),
attempts to gain physical insight have been based so far on
analytical models that {\em assume} the existence of power-law XLFs, and that X-ray 
lifetime is  inversely proportional to 
X-ray luminosity and XRB properties do not depend on the SF history of the host 
galaxy (Wu 2001; Kilgard et al. 2002). 

In order to develop a set of useful diagnostics, it is 
important to develop theoretical models for XRB formation and evolution that depend 
on the star-formation history of the host galaxies, and allow us to identify the 
main physical elements that determine the XLF slopes. Given that this study focuses on 
observed {\em populations}, it is clear that the employment of population synthesis 
models is necessary. 
First pre-Chandra XRB models for starbursts were developed and compared to ASCA 
observations of the total X-ray luminosity of WR
galaxy He2-10 by Van Bever \& Vanbeveren 2000.
However, only now we may compare the specific theoretical models with the
observed point source populations.  
A collaborative effort has led to the development of a detailed    
population synthesis code {\em StarTrack} (see \S\,\ref{sec:models}) 
that specifically focus on careful, self-consistent XRB calculations.
Ultimately our goal is to construct a coherent picture of XRB 
formation and evolution based on 
theoretical models that have been first calibrated against observations of well-studied 
galaxies, and thus can be used in the interpretation of other XLF observations.

In this {\em Letter} we compare our model XLFs to {\em Chandra} observations of 
NGC~1569, a blue dwarf Irregular at a distance of 2.2Mpc (Israel 1988), characterized 
as a (post-)starburst galaxy. It has been selected as a good test case because its star 
formation history is {\em relatively} well constrained by HST optical and infrared 
observations (Aloisi et al. 2001; Greggio et al.1998; Vallenari \& Bomans 1996) 
{\em and} has a long ($\sim 80$\,ks) {\em Chandra} exposure providing a detection limit of 
$\simeq 10^{36}$\,erg\,s$^{-1}$ (Martin, Kobulnicky \& Heckman 2002; hereafter M02). 
We are mainly concerned with two questions: (i) is it at all possible to theoretically 
reproduce the observed XLF? and  (ii) do our models agree with the current constraints 
on the star-formation history of NGC~1569 derived by observations in other wavelenghts?

\section{THEORETICAL MODELS}
 \label{sec:models}

The {\em StarTrack} code was originally developed for the modeling of binaries with 
two compacts (BKB), but has recently undergone major revisions 
(Belczynski et al. 2003, in preparation) 
intended to treat in detail the formation and evolution of XRBs, for any choice of 
SF history and metallicity. The main revisions include a 
detailed treatment of tidal synchronization and circularization (Hut 1981), 
individual treatment of various mass-transfer (MT) episodes, 
full numerical orbit evolution with angular momentum losses due to magnetic breaking, 
gravitational radiation, mass transfer/loss, and tides. 

We have calibrated the tidal implementation using observations of binary 
eccentricities and periods in stellar clusters (Mathieu et al.\ 1992) and of orbital 
decay in high-mass XRBs  (Levine et al.\ 2000). 
Our MT implementation involves the detailed 
calculation of Roche-lobe overflow MT rates based on radius-mass exponents 
 both for the donor stars and their Roche lobes. We have compared our results 
for a set of MT sequences to both published MT calculations (e.g., Beer \& 
Podsiadlowski 2002) and results 
obtained within our group using  an updated stellar evolution code (Ivanova et al. 2003) 
with very satisfactory agreement, much better than typical MT implementations. 
The modeled X-ray phases are 
identified as (i) stable, driven by nuclear evolution of the donor, magnetic braking 
or gravitational radiation 
(ii) thermally unstable with possibly anisotropic emission (King et al. 2001); 
(iii) Eddington-limited MT (cf.\ Kalogera et al.\ 2003, in preparation); and 
(iv) persistent or transient (critical MT rate from Dubus et al. 1999). We also 
account for wind 
accretion onto compact objects following Hurley, Tout \& Pols (2002). 
The updated code has been already tested and used for a study of 
Galactic ultracompact binaries (Belczynski \& Taam 2003).

In this paper we examine whether XRB models with rather standard binary-evolution 
parameters (i.e., not specifically selected for this study), but with star-formation 
history and metallicity consistent with what is known about NGC~1569 can produce an XLF 
shape 
in agreement with observations. We choose parameters from the reference model  in BKB, 
with just a few differences: the maximum neutron star (NS) mass is set equal to $2 \msun$, 
the most 
recently inferred natal NS kick distribution is incorporated (Arzoumanian, Chernoff \& 
Cordes 2002), the primary 
masses are selected in the range $4-100 \msun$ (with an initial-mass-function slope of 
-2.7; see Kroupa, Tout \& Gilmore 1993), and the secondary masses in the range 
$0.08-100 \msun$ with a flat mass ratio (secondary divided by primary mass) distribution. 

We construct our XRB models with focus on the estimated metallicities and
SFR properties of stellar populations in NGC~1569. A coherent picture has been 
developed (Aloisi et al. 2001; Greggio et al. 1998; Vallenari \& Bomans 1996) for two 
populations: (i) one 
metal-rich~($\sim 0.25$\,Z$_\odot$, although it could be comparable to 1\,Z$_\odot$; M02), 
young population formed in a global burst of star formation that lasted about 100 Myr and 
seems to have stopped 5-10Myr ago, and (ii) one metal-poor ($Z=0.004-0.0004$), old 
population formed by less vigorous (possibly by a factor of $10-20$ in SFR), continuous 
star formation for the past 1-1.5 Gyr. For consistency with these estimates we adopt 
$Z=0.005$ and $Z=0.0022$, for the young and the old populations, respectively. 
We examine the consistency of our models by allowing the relative SFR
weight and the population age and metallicity  to vary.

\section{THE X-RAY LUMINOSITY FUNCTION} 
 \label{sec:results} 
 
We analyzed the archival {\em Chandra} data using the Ciao 
v3.0 data analysis suite. After screening for high background intervals we
searched for sources in the full 0.3-7.0~keV band with the
{\textit{wavedetect}} source detection algorithm. Then, for each source
within the D25 ellipse of the galaxy ($\rm{D_{max}=3.6'}$, $\rm{D_{min}=1.8'}$), 
we estimated the net 
number of counts using an aperture including all the
emission from a source. The background was determined locally from a
source-free annulus around each source. We detected 14 sources 
with a significance greater than 3$\sigma$ above the local background, in excellent 
agreement with M02, but we exclude one that has been identified as a supernova 
remnant by M02. Based on the $\textit{LogN-LogS}$ relation from the $\textit{Champ}$
survey (Kim et al.\ 2003) we estimate that at most two of our sources
within the D25 area are associated with background or foreground objects.

The relatively high number of counts of the faintest sources 
($\sim15-20$ counts) compared to their typical background ($\sim5-10$ counts) and the 
lack of a turnover in the observed cumulative XLF (see Figure 2)  
indicate that incompleteness is not significant in the low luminosity end of the XLF. 
The higher level of diffuse emission in the central region of the galaxy (M02) may result 
to a higher detection threshold, but, given the very small number of sources in this region, 
we do not attempt to correct for this effect. In our XLF, apart from the standard 
Gehrels errors (Gehrels 1986), we include errors associated with the uncertainties in 
the observed count rate of each source, following Zezas \& Fabbiano (2002). The 
luminosity of the sources is in the 0.1-10.0~keV band assuming a power-law spectrum 
($\Gamma=1.7$) with Galactic column density ($2.3\times10^{20}~\rm{cm^{-3}}$) and is 
corrected for absorption.
   
In our theoretical models X-ray luminosities ($L_X$) are calculated based on the 
{\em accretion} rate $\dot{M}$ with appropriate efficiencies for 
NS and black holes (BH). At present no corrections specific to the {\em Chandra} 
energy band are taken into account, due to the lack of a 
reliable spectral model across a wide range of frequencies. We further apply the Eddington 
limit and calculate the effects of associated non-conservative mass transfer with the lost 
matter carrying away the specific angular momentum of the accretor.   

To determine the contribution of transient systems to the XLF we need information on their 
X-ray duty cycle (DC), which however cannot be provided reliably by the disk instability 
theory.  Empirically it is thought that $DC \lesssim 1\%$ (e.g., Taam, King \& Ritter 
2000). In {\em StarTrack} we adopt a DC value (in this study DC=1\%) and, for each 
system we randomly sample the probability that the source is
in outburst and then assign an X-ray luminosity equal to the Eddington limit,
whereas for the systems in quiescence we assume that their X-ray luminosities 
are too low to be detectable.
Donors more massive than the accretors can drive  mass transfer on their thermal 
timescale (e.g., Kalogera \& Webbink1996) and anisotropic emission is 
possible (King et al. 2001). We identify such systems based on the calculated radius-mass 
exponents. At present we do not account for the anisotropic-emission possibility, but 
do account for the possibility of super-Eddington transfer as described above. 
Both transient and thermal-timescale XRBs are Roche-lobe overflow (RLOF) systems,
and typically dominate old ($\gtrsim 100$\,Myr) populations (see also Sipior 
et al.\ 2003).

\begin{figure}
\begin{center}
\psfig{figure=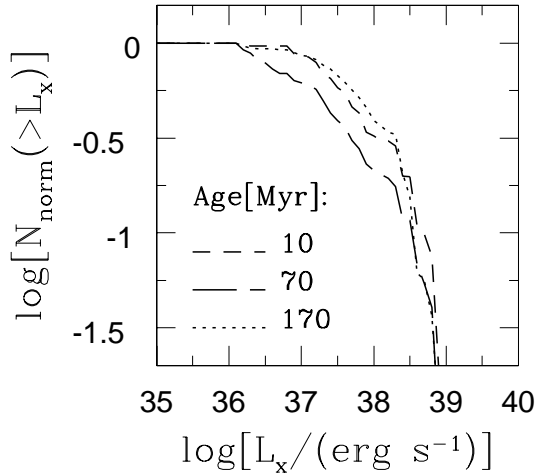,width=2.85in}
\caption{Non-monotonic behavior with time of theoretical normalized XLFs for 
two stellar populations: one old at 1.5 Gyr  and one young at age 10, 70, and 
170\,Myrs (continuous SFR through 1.5\,Gyr, and 10, 70, and 100\,Myrs, 
respectively). the average SFR in the old population is assumed to be 20 times 
smaller than that in the young population.}
\vspace*{-1.3cm}
\label{fig:mod00}
\end{center}
\end{figure}

\begin{figure}
\begin{center}
\psfig{figure=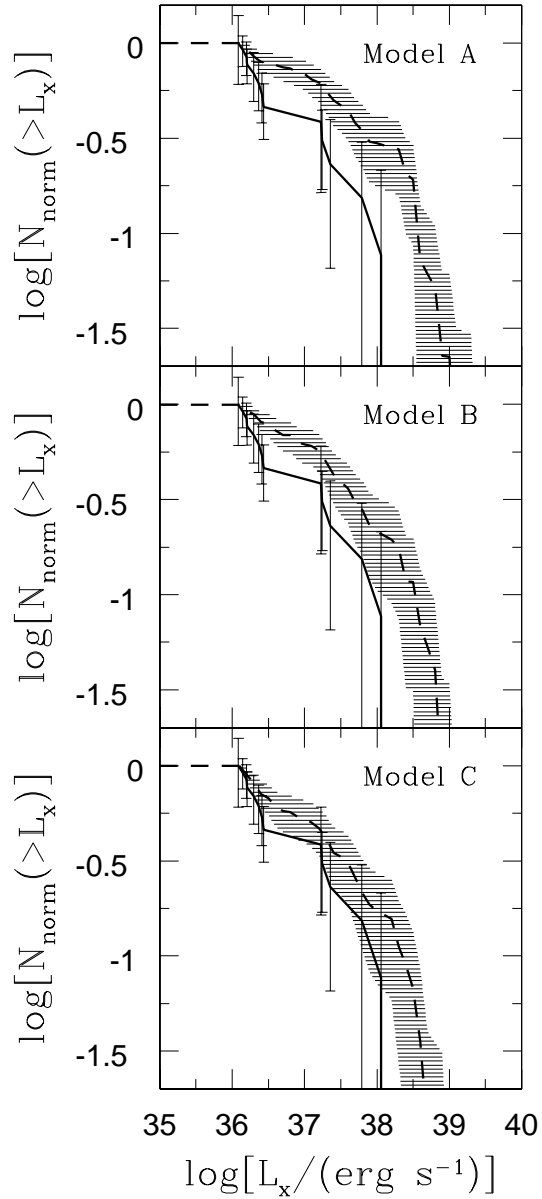,width=2.85in}
\caption{The observed NGC1569 XLF with error bars (solid lines) plotted against 
predicted XLFs (dashed curves) shown with $1\,\sigma$\ sampling errors
(shaded areas).  All curves are normalized to the total number (13) of the 
detected sources. Each panel corresponds to different choices for the age 
of the young and old populations and the SFR ratio of the young relative to 
the old, respectively.  
{\em Top panel:} 110\,Myr, 1.5\,Gyr, and 20;     
{\em Middle panel:} 70\,Myr, 1.5\,Gyr, and 20;  
{\em Bottom panel:} 70\,Myr, 1.3\,Gyr, and 40.}
\vspace*{-1.3cm}
\label{fig:mod01}
\end{center}
\end{figure}

Based on the current understanding of the SF history of NGC1569 
(\S\,\ref{sec:models}), we calculate combined XLFs for ages 1--1.5\,Gyr and 
70--110\,Myr, for the old and young populations, respectively. 
The choice of the relative weight of the two populations also affects the XLF 
shape and we choose SFR weight factors in the range 10--50. Population 
models tend to have rather uncertain absolute normalizations and the absolute
SFRs in NGC~1569 are not well constrained either (Vallenari \& Bomans1996). 
Although, the total number 
of sources in a galaxy (corrected carefully for selection effects, partial galaxy 
coverage, 
etc) can provide additional model constraints, in this first step of our studies we 
restrict ourselves to calibrations and comparisons to observations based on the XLF 
{\em shape}. Therefore we used observed and model XLFs normalized to the total number 
of objects, and we take into account the small-number bias by randomly selecting 
from the simulated source populations samples of 13 sources at a time. 
The sampling of the
underlaying model parent population is then repeated number of times ($\sim 10^4$) to 
yield the predicted XLFs and their associated errors.

Figure 1 indicates the behavior of XLF as a function of age of young XRB population 
of NGC1569, while the age of the old population is fixed at 1.5\,Gyr.
Unlike earlier suggestions (Wu 2001; Kilgard et al. 2002), we find that the 
dependence of the 
XLF slope on age is {\em non-monotonic} and this can be actually understood physically. 
At early times ($\sim 10$\,Myr) the XLFs are by far dominated by MT-fed
systems of the older population. The recent star formation episode barely begun
producing wind-fed systems with the most massive compact objects (formed earlier)
that accrete from strong winds from the next most massive donors. 
The combination of young and bright wind-fed sources with the MT-fed systems of
older population tends to make XLFs rather flat.
However, relatively quickly ($\sim 70$\,Myr) the very massive stars of young population 
end their lives 
and are removed from the XRB sample. They are replaced by less massive, and fainter 
wind-fed systems and the XLF becomes steeper. 
As time passes ($\sim 170$\,Myr) RLOF systems of younger population, that tend to be 
brighter than wind-fed sources of similar binary component masses, start becoming 
important in the total population and the XLF slope once again becomes flatter.

In Figure 2 we present XLFs calculated for NGC1569. 
Model A with parameters taken at face 
value from the HST observations: young population at 110\,Myrs 
with SFR of 20 times larger than that of the old population at 1.5\,Gyr.
The calculated function tends to be flatter and extends to higher $L_X$ 
values compared to the observed XLF. 
At the age of 110\,Myr since the onset of the burst (10\,Myr after its end), 
the young population includes a significant number of a bright RLOF sources 
flattening the XLF. 
Model B corresponds to the young population at an earlier age of 70\,Myrs 
(equal to the burst duration)
when it is dominated by high mass XRBs, and we already see that 
the XLF becomes steeper, but not quite enough to match the observed one. 
There are still quite a few RLOF systems formed in the old star formation 
episode. However, the duration of that episode as well as its end time  
are not very precisely established.
Model C corresponds to the young population at 
70\,Myrs (equal to the burst duration), but the old one at 1.3\,Gyr with continuous 
star formation for 1\,Gyr. 
The predicted XLF quite closely match the observed one within the relevant errors. 
We note the tendency for higher luminosities in the models, but we note that we 
adopt bolometric luminosities, which are higher than those in the {\em Chandra} 
band.
In Table 1 we present the content of the X-ray binary population
for our best model (C). The population is dominated 
by young high-mass XRBs, but with significant contribution of 
old RLOF systems. In the young population, where only the most massive 
stars have ended evolution, many accretors are BH, while in the 
older population sources with NS dominate.

\begin{deluxetable}{crr}
\tablewidth{500pt}
\tablecaption{Model C XRB Sub-Populations\tablenotemark{a}}
\tablehead{ Type                        & Old Pop.     & Young Pop. }
\startdata
RLOF systems                            & 24.4\%       & 3.7\%     \\
NS/BH accretor                          &  1.5         & 4.0       \\
transient\tablenotemark{b}/persistent   &  0.1         & no trans. \\
thermal/rest                            &  0.38        & 0.25      \\
                                        &              &           \\
WIND-fed systems                        &  1.5\%       & 70.4\%    \\
NS/BH accretor                          &  1.0         & 0.53      

\enddata
\label{numbers01}
\tablenotetext{a}{Corresponds to all active ($L_x>1.22 \times 10^{36} erg\ s^{-1}$) 
sources.} 
\tablenotetext{b}{Only transients at outburst stage are listed here assuming DC=1\%.}
\end{deluxetable}

\section{DISCUSSION} 
\label{sec:discussion} 
 
We present our first results from XRB population models developed for comparison 
with current and future {\em Chandra} observations of nearby galaxies. 
We choose NGC~1569 as our first test case and we find good agreement
between our models and the observations. This agreement is even more 
remarkable in view of the fact that we did not
attempt to fine tune any of the model parameters related to X-ray binary 
evolution. 
However, we have explored other models with varying metallicities 
and IMF slopes, and found that both our quantitative and qualitative conclusions 
remain robust  and the other models do not
offer a better match to the observed XLF shape. This is true even when we account for 
the fraction of systems that can escape NGC~1569, due to systemic
velocities acquired at supernova explosions.

Examination of various models with properties consistent with NGC~1569 
constraints, lead us to conclusion that an age of 70\,Myr for the young
and 1.3 Gyr for the old population and a SFR relative weight of 40 are 
favored.
In order to get agreement between the model and the observed 
XLF, we require a recent burst that is younger than inferred 
from the optical/infrared data.
This slight discrepancy could be due to the fact that at this point we do not 
consider different black-hole binary spectral states and anisotropic emission 
from pulsar binaries. On the other hand the parameters of the older population
(which is dominated by old non-magnetized neutron star binaries) are very 
consistent with the latest picture from the HST data (Angeretti et al.\ 2003, 
private communication).

We consider these encouraging results only a small, first step in our exploration of XRB 
models and their comparison to observations. As we gain experience with the 
study of specific galaxies, we expect to develop a reliable calibration system that will 
then allow us to extract information about origin of XRBs in other galaxies. A natural 
extension of this study will include two elements: the exploration of constraints on the 
absolute normalization of the XLF in addition to its shape, and the comparison models 
with a sample of starburst galaxies that form a time sequence with ages in a wide range 
of values to address the theoretical basis for correlations suggested by Grimm et al. 
(2003), for example. Detailed examination of degeneracies in the derived constraints is 
also important.  
Moreover, the modeling of supernovae remnants may prove to
be necessary, since they may {\em i)} contribute significantly to and {\em
ii)} be hard
to remove from observed point source samples.

\acknowledgments
We thank the referee J.~Irwin, and also A.~King and 
T.~Maccarone for useful comments, and the Aspen Center for Physics and the 
NU Visitors' fund (to AZ) for support. 
We also thank M.~Tosi and L.~Angeretti for discussing their results prior to 
publication.
This work is partially supported by a Packard Fellowship, and 
a {\em Chandra} theory grant to VK, NASA LTSA grant NAG5-13056 to AZ and VK
and NASA grant NAS8-39073 to GF.


\begin{references}

\reference{} Aloisi et al.\ 2001, \apj , 121,1425
\reference{} Arzoumanian, Z., Chernoff, D.\ F., \& Cordes, 
             J.\ M.\ 2002, \apj, 568, 289
\reference{} Beer, M.E.\ \& Podsiadlowski, P.\ 2002, \mnras , 331, 351
\reference{} Belczynski, K., Kalogera, V., \& Bulik, T.\ 2002,
             \apj, 572, 407 (BKB)
\reference{} Belczynski, K., \& Taam, R.E.\ 2003, \apj, submitted
\reference{} Dubus, G.\ et al.\ 1999, \mnras , 303, 139
\reference{} Fabbiano, G.\ \& White, N.\ 2003, review 
\reference{} Gehrels, N.\ 1986, \apj, 303, 336 
\reference{} Greggio, L., et al.\ 1998, \apj , 504, 725
\reference{} Grimm, H.-J, Gilfanov, M., \& Sunyaev, R.\ 2003, \mnras , 339, 793
\reference{} Hurley, J. R., Tout, C. A., \& Pols, O. R.\ 2002, \mnras, 329, 897
\reference{} Hut, P.\ 1981, \aap, 99, 126 
\reference{} Israel, F. P.\ 1988, \aap, 194, 24
\reference{} Ivanova, N., Belczynski, K., Kalogera, V., Rasio, F., \& 
             Taam, R. E.\ 2003, \apj, 592, 475
\reference{} Kalogera, V.\ \& Webbink, R.F.\ 1996, \apj , 458, 301
\reference{} Kilgard, R.E., et al.\ 2002, \apj , 573, 138
\reference{} Kim et al.\ 2003, \apj, in press (astro-ph/0308493)
\reference{} King, A.R.\ et al.\ 2001, \apj , 552, L109
\reference{} Kroupa, P., Tout, C.\ A., \& Gilmore, G.\ 1993,
             \mnras, 262, 545
\reference{} Levine, A., Rappaport, S.A., \& Zojcheski, G.\ 2000, 541, L194
\reference{} Martin, C.~L., Kobulnicky, H.~A., \& Heckman, T.~M.\ 2002, \apj, 
             574, 663 (M02) 
\reference{} Mathieu, R. D., et al.\ 1992, "Binaries as Tracers of 
             Stellar Formation", ed. Duquennoy, A. \& Mayor, M., Cambridge 
             University Press, p.278 
\reference{} Sarazin, C.L., et al.\  2003, \apj , 595, in press 
\reference{} Sipior, M., Eracleous, M., \& Sigurdsson, S.\ 2003, \apj , submitted 
             (astro-ph/0308077)
\reference{} Taam, R.E., King, A.R., \& Ritter H.\ 2000, \apj, 541, 329
\reference{} Vallenari, A.\ \& Bomans, D.J.\ 1996, \aap , 313, 713
\reference{} Van Bever, J., \& Vanbeveren, D.\ 2000, \aap, 358, 462
\reference{} Wu, K.\ 2001, Publications of the Astronomical Society of Australia, 18, 443
\reference{} Zezas, A.\ \& Fabbiano, G.\ 2002, \apj , 577, 726

\end{references}
\end{document}